
\input amssym.tex
\input epsf
\input pstricks

\def\cS{{\cal S}}

\def\by{{\bf y}}
\def\bz{{\bf z}}


\magnification=\magstephalf
\hsize=16.0 true cm
\vsize=22.0 true cm
\hoffset=0.0 true cm
\voffset=1.0 true cm

\abovedisplayskip=12pt plus 3pt minus 3pt
\belowdisplayskip=12pt plus 3pt minus 3pt
\parindent=0.0em


\font\sixrm=cmr6
\font\eightrm=cmr8
\font\ninerm=cmr9

\font\sixi=cmmi6
\font\eighti=cmmi8
\font\ninei=cmmi9

\font\sixsy=cmsy6
\font\eightsy=cmsy8
\font\ninesy=cmsy9

\font\sixbf=cmbx6
\font\eightbf=cmbx8
\font\ninebf=cmbx9

\font\eightit=cmti8
\font\nineit=cmti9

\font\eightsl=cmsl8
\font\ninesl=cmsl9

\font\sixss=cmss8 at 8 true pt
\font\sevenss=cmss9 at 9 true pt
\font\eightss=cmss8
\font\niness=cmss9
\font\tenss=cmss10

 at 12 true pt
\font\bigrm=cmr10 at 12 true pt
 at 12 true pt

\font\Bigbss=cmssbx12 at 16 true pt
 at 14 true pt
 at 14 true pt
 at 16 true pt
 at 14 true pt

\font\teneufm=eufm10
\font\seveneufm=eufm7
\font\fiveeufm=eufm5
\newfam\eufmfam
\textfont\eufmfam=\teneufm
\scriptfont\eufmfam=\seveneufm
\scriptscriptfont\eufmfam=\fiveeufm
\def\eufm#1{{\fam\eufmfam\relax#1}}

\font\tencmmib=cmmib10 \skewchar\tencmmib='177
\newfam\cmmibfam
\textfont\cmmibfam=\tencmmib

\catcode`@=11
\newfam\ssfam

\def\tenpoint{\def\rm{\fam0\tenrm}%
    \textfont0=\tenrm \scriptfont0=\sevenrm \scriptscriptfont0=\fiverm
    \textfont1=\teni  \scriptfont1=\seveni  \scriptscriptfont1=\fivei
    \textfont2=\tensy \scriptfont2=\sevensy \scriptscriptfont2=\fivesy
    \textfont3=\tenex \scriptfont3=\tenex   \scriptscriptfont3=\tenex
    \textfont\itfam=\tenit                  \def\it{\fam\itfam\tenit}%
    \textfont\slfam=\tensl                  \def\sl{\fam\slfam\tensl}%
    \textfont\bffam=\tenbf \scriptfont\bffam=\sevenbf
    \scriptscriptfont\bffam=\fivebf
                                            \def\bf{\fam\bffam\tenbf}%
    \textfont\ssfam=\tenss \scriptfont\ssfam=\sevenss
    \scriptscriptfont\ssfam=\sevenss
                                            \def\ss{\fam\ssfam\tenss}%
    \normalbaselineskip=19pt
    \setbox\strutbox=\hbox{\vrule height8.5pt depth3.5pt width0pt}%
    \let\big=\tenbig
    \normalbaselines\rm}

\def\ninepoint{\def\rm{\fam0\ninerm}%
    \textfont0=\ninerm      \scriptfont0=\sixrm
                            \scriptscriptfont0=\fiverm
    \textfont1=\ninei       \scriptfont1=\sixi
                            \scriptscriptfont1=\fivei
    \textfont2=\ninesy      \scriptfont2=\sixsy
                            \scriptscriptfont2=\fivesy
    \textfont3=\tenex       \scriptfont3=\tenex
                            \scriptscriptfont3=\tenex
    \textfont\itfam=\nineit \def\it{\fam\itfam\nineit}%
    \textfont\slfam=\ninesl \def\sl{\fam\slfam\ninesl}%
    \textfont\bffam=\ninebf \scriptfont\bffam=\sixbf
                            \scriptscriptfont\bffam=\fivebf
                            \def\bf{\fam\bffam\ninebf}%
    \textfont\ssfam=\niness \scriptfont\ssfam=\sixss
                            \scriptscriptfont\ssfam=\sixss
                            \def\ss{\fam\ssfam\niness}%
    \normalbaselineskip=16pt
    \setbox\strutbox=\hbox{\vrule height8.0pt depth3.0pt width0pt}%
    \let\big=\ninebig
    \normalbaselines\rm}

\def\eightpoint{\def\rm{\fam0\eightrm}%
    \textfont0=\eightrm      \scriptfont0=\sixrm
                             \scriptscriptfont0=\fiverm
    \textfont1=\eighti       \scriptfont1=\sixi
                             \scriptscriptfont1=\fivei
    \textfont2=\eightsy      \scriptfont2=\sixsy
                             \scriptscriptfont2=\fivesy
    \textfont3=\tenex        \scriptfont3=\tenex
                             \scriptscriptfont3=\tenex
    \textfont\itfam=\eightit \def\it{\fam\itfam\eightit}%
    \textfont\slfam=\eightsl \def\sl{\fam\slfam\eightsl}%
    \textfont\bffam=\eightbf \scriptfont\bffam=\sixbf
                             \scriptscriptfont\bffam=\fivebf
                             \def\bf{\fam\bffam\eightbf}%
    \textfont\ssfam=\eightss \scriptfont\ssfam=\sixss
                             \scriptscriptfont\ssfam=\sixss
                             \def\ss{\fam\ssfam\eightss}%
    \normalbaselineskip=10pt
    \setbox\strutbox=\hbox{\vrule height7.0pt depth2.0pt width0pt}%
    \let\big=\eightbig
    \normalbaselines\rm}

\def\tenbig#1{{\hbox{$\left#1\vbox to8.5pt{}\right.\n@space$}}}
\def\ninebig#1{{\hbox{$\textfont0=\tenrm\textfont2=\tensy
                       \left#1\vbox to7.25pt{}\right.\n@space$}}}
\def\eightbig#1{{\hbox{$\textfont0=\ninerm\textfont2=\ninesy
                       \left#1\vbox to6.5pt{}\right.\n@space$}}}

\font\sectionfont=cmbx10
\font\subsectionfont=cmti10

\def\figurecaptionfont{\ninepoint}
\def\tablecaptionfont{\ninepoint}


\newcount\equationno
\newcount\bibitemno
\newcount\figureno
\newcount\tableno

\equationno=0
\bibitemno=0
\figureno=0
\tableno=0


\footline={\ifnum\pageno=0{\hfil}\else
{\hss\rm\the\pageno\hss}\fi}


\def\section #1. #2 \par
{\vskip0pt plus .20\vsize\penalty-100 \vskip0pt plus-.20\vsize
\vskip 1.4 true cm plus 0.2 true cm minus 0.2 true cm
\global\def\equationlabel{#1}
\global\equationno=0
\leftline{\sectionfont #1. #2}\par
\immediate\write\terminal{Section #1. #2}
\vskip 0.4 true cm plus 0.1 true cm minus 0.1 true cm
\noindent}


\def\subsection #1 \par
{\vskip0pt plus 0.4 true cm\penalty-50 \vskip0pt plus-0.4 true cm
\vskip2.5ex plus 0.1ex minus 0.1ex
\leftline{\subsectionfont #1}\par
\immediate\write\terminal{Subsection #1}
\vskip1.0ex plus 0.1ex minus 0.1ex
\noindent}


\def\appendix #1 #2 \par
{\vskip0pt plus .20\vsize\penalty-100 \vskip0pt plus-.20\vsize
\vskip 1.6 true cm plus 0.2 true cm minus 0.2 true cm
\global\def\equationlabel{\hbox{\rm#1}}
\global\equationno=0
\leftline{\sectionfont Appendix #1 \quad {\it #2}}\par
\immediate\write\terminal{Appendix #1 #2}
\vskip 0.7 true cm plus 0.1 true cm minus 0.1 true cm
\noindent}



\def\equation#1{$$\displaylines{\qquad #1}$$}
\def\enum{\global\advance\equationno by 1
\hfill\llap{(\equationlabel.\the\equationno)}}


\def\ifundefined#1{\expandafter\ifx\csname#1\endcsname\relax}

\def\ref#1{\ifundefined{#1}?\immediate\write\terminal{unknown reference
on page \the\pageno}\else\csname#1\endcsname\fi}

\newwrite\terminal
\newwrite\bibitemlist

\def\bibitem#1#2\par{\global\advance\bibitemno by 1
\immediate\write\bibitemlist{\string\def
\expandafter\string\csname#1\endcsname
{\the\bibitemno}}
\item{[\the\bibitemno]}#2\par}

\def\beginbibliography{
\vskip0pt plus .15\vsize\penalty-100 \vskip0pt plus-.15\vsize
\vskip 1.2 true cm plus 0.2 true cm minus 0.2 true cm
\leftline{\sectionfont References}\par
\immediate\write\terminal{References}
\immediate\openout\bibitemlist=biblist
\frenchspacing\parindent=1.8em
\vskip 0.5 true cm plus 0.1 true cm minus 0.1 true cm}

\def\endbibliography{
\immediate\closeout\bibitemlist
\nonfrenchspacing\parindent=1.0em}

%
%
\def\figurecaption#1{\global\advance\figureno by 1
\narrower\figurecaptionfont
Fig.~\the\figureno. #1}
\def\tablecaption#1{\global\advance\tableno by 1
\vbox to 0.5 true cm { }
\centerline{\tablecaptionfont%
Table~\the\tableno. #1}
\vskip-0.4 true cm}
\def\thicktablerule{\hrule height1pt}
\def\thintablerule{\hrule height0.4pt}
\tenpoint

\immediate\openin\bibitemlist=biblist
\ifeof\bibitemlist\immediate\closein\bibitemlist
\else\immediate\closein\bibitemlist
\input biblist \fi
%
%

\def\su3id{{\Bbb I}}

%
%
\def\lvec#1{\setbox0=\hbox{$#1$}
    \setbox1=\hbox{$\scriptstyle\leftarrow$}
    #1\kern-\wd0\smash{
    \raise\ht0\hbox{$\raise1pt\hbox{$\scriptstyle\leftarrow$}$}}
    \kern-\wd1\kern\wd0}
\def\rvec#1{\setbox0=\hbox{$#1$}
    \setbox1=\hbox{$\scriptstyle\rightarrow$}
    #1\kern-\wd0\smash{
    \raise\ht0\hbox{$\raise1pt\hbox{$\scriptstyle\rightarrow$}$}}
    \kern-\wd1\kern\wd0}
%
%

%
%
\def\psibar{\overline{\psi}}

\def\zetabar{\overline{\zeta}}
%
%
\def\euf{{\eufm f}}
%
%

\def\btheta{\hbox{$\mathchar"0\hexnumber@\cmmibfam12$}}
%
%
\def\NPB #1 #2 #3 {Nucl.~Phys.~{\bf#1} (#2)\ #3}
\def\NPBproc #1 #2 #3 {Nucl.~Phys.~B (Proc. Suppl.) {\bf#1} (#2)\ #3}
\def\PRD #1 #2 #3 {Phys.~Rev.~{\bf#1} (#2)\ #3}
\def\PLB #1 #2 #3 {Phys.~Lett.~{\bf#1} (#2)\ #3}
\def\PRL #1 #2 #3 {Phys.~Rev.~Lett.~{\bf#1} (#2)\ #3}
\def\PR  #1 #2 #3 {Phys.~Rep.~{\bf#1} (#2)\ #3}
\def\CPC #1 #2 #3 {Comput.~Phys.~Commun.\  {\bf #1} (#2) #3}
%

\def\fBsmall{483(4)}
\def\fB{170(11)(5)(22)}
\def\fBstat{170(11)}

\def\fBssmall{490(4)}
\def\fBs{192(9)(5)(24)}

\def\fD{204(9)}
\def\fDsmall{634(6)}

\def\fBsd{1.13(2)(1)}
\def\fDsd{1.10(1)(1)}

%
\pageno=0
\hfill 
\vskip 3.3 true cm 
\centerline
{\Bigbss \scalebox{1.4 1.4}{${\bf f_B}$} and two scales problems in lattice QCD}
\vskip 1.0 true cm
\centerline{\bigrm  Marco Guagnelli, Filippo Palombi, Roberto Petronzio, Nazario Tantalo\kern1pt}
\vskip 0.8 true cm
\centerline{\it  Dipartimento di Fisica, Universit\`a di Roma Tor Vergata}
\vskip 0.15 true cm
\centerline{\it and INFN, Sezione di RomaII,}
\vskip 0.3 true cm
\centerline{\it Via della Ricerca Scientifica 1, 00133 Rome, Italy} 
\vskip 3.0 true cm
\thintablerule
\vskip 2.0ex
\ninepoint
\leftline{\bf Abstract} 
\vskip 1.0ex\noindent
A novel method to calculate $f_B$ on the lattice is introduced, based on the study of
the dependence of finite size effects upon the heavy quark mass of flavoured mesons and on a non--perturbative 
recursive finite size technique. We avoid the systematic errors related to extrapolations
from the static limit  or to the tuning of the coefficients of effective Lagrangian and
the results admit an extrapolation to the continuum limit.
We perform a first estimate at finite lattice spacing, but close to the continuum limit, giving 
$f_B = \fB$ {\rm MeV}. We also obtain $f_{B_s} = \fBs${\rm MeV}. The first error is statistical, 
the second is our estimate of the systematic error from the method and the third the 
systematic error from the specific approximations adopted in this first exploratory calculation.
The method can be generalized to two--scale problems in 
lattice QCD.
\vskip 0.3cm
Keywords: lattice QCD, decay constants, $B$ meson.

\vskip 2.0ex
\thintablerule
\vskip 8.0ex 
\centerline{September 2002}
\vfill\eject
\footline={\ifnum\pageno=0{\hfil}\else
{\hss\rm\the\pageno\hss}\fi}
\tenpoint
\section 1. Introduction

Lattice QCD evaluations of quantities characterised by two scales with a large
hierarchy require in general a very high lattice resolution and a sizeable total 
physical volume to correctly account the dynamics of the small distance  scale and to
dispose of the finite size effects related to the large distance  scale. A good example
is provided by the pseudoscalar $B$  meson decay constant~[\ref{litone}],  where the small
distance scale is represented by the inverse of the bottom quark mass and
the large distance scale by the radius of the $B$ meson, related in turn to the inverse of the light quark mass.
A straight evaluation of the decay constant would require lattices with $N=80^4$ points or more,
 exceeding the present generation computers capabilities, and, in the case
of unquenched simulations, the ones of the next generation.
One resorts to approximate calculations based on extrapolations from the static limit
or on non--relativistic formulations of standard QCD. All the available methods introduce
systematic errors related to extrapolation fits and/or to the use of effective Lagrangians. 
We present a novel approach based on the study of the
dependence upon the heavy quark mass of finite size effects for the pseudoscalar decay constant
of heavy flavoured mesons. The basic assumption is that the finite size effects are mainly related 
to the light quark mass  and rather insensitive to the one of a sufficiently heavy quark.
We discuss the general features of the method assuming the continuum limit has been taken.
Corrections specific to the finite lattice spacing calculation presented in this first paper are discussed later. 
The relevant quantity is the ratio $\sigma$ of the pseudoscalar constants at
different volumes:
\equation{
\sigma\ \equiv\ {\ f_B( 2L)\ \over f_B(L)}
\enum}

where $f_B(L)$ is the value of the decay constant on a volume with linear size $L$.
The dimensionless $\sigma$ depends on general grounds upon three dimensionless
variables: $m_\ell L$, $m_hL$ and $\Lambda_{QCD}L$.  
For a sufficiently large heavy quark mass $m_h$, the dependence is basically dominated by the light quark 
and the expansion for large $m_h$ takes the form
\equation{
\sigma \ = \ \sigma\bigl( m_\ell L,\ \Lambda_{QCD}L\bigr)\ +\ {C\bigl(m_\ell L,\ \Lambda_{QCD}L\bigr)\over{m_h L}}
\enum}

A simple phenomenological ansatz for $\sigma$ can be made based on the concept of
a reduced mass constructed out of the heavy and light quark masses
\equation{
\sigma = \sigma\bigl(m_{red} L,\ \Lambda_{QCD}L\bigr)
\enum}

where 
\equation{
m_{red} = {\mu_1 \mu_2 \over \mu_1 + \mu_2}
\enum}  

The quantity $\mu_i$ is a function of the quark mass, but
not only: indeed, for very light masses, finite size
effects are regulated by the physical meson size, which is expected to remain finite when the
light quark mass tends to zero. We will show later some evidence for $\mu_i$ being a simple linear
combination of the light quark mass and $\Lambda_{QCD}$.
\vskip 0.5cm
A crucial question is the threshold value of the quark mass on a given volume where the
large $m_h$ expansion becomes reliable. As we will see, this value falls in a mass range of the
order of a couple of GeV in the renormalization invariant mass scheme, where the calculation on a
single lattice is affordable.
Under these circumstances, the strategy to obtain $f_B$ is the following. One first
performs a calculation on a lattice where the resolution is suitable for $b$ quark
propagation, but the total volume is unavoidably a small one. This sets $f_B$ on a finite volume.
In order to connect to the large volume results, one needs the step scaling function $\sigma$ 
for values of heavy quark masses generally lower  than those of the simulation where the
finite size value of $f_B$ was obtained.
The possibility of extrapolating $\sigma$ to heavier masses depends upon the validity of the
asymptotic expansion: in a favourable case, as will be the real one, one can evaluate the finite
size effects in a reliable way, connecting, by a repeated iteration of the procedure, small volume values of $f_B$
to the ones on large volumes,
\equation{
f_B^{phys} \ = \ f_B(L_0)\ \sigma(L_0)\ \sigma(2L_0)\ \dots
\enum} 

and the recursion stops on a volume where $\sigma \simeq 1$ within a required precision.
The continuum limit is obtained by extrapolating to zero lattice spacing the step scaling function
obtained at fixed physical quantities. 
This paper deals with a first exploration of the method at finite lattice spacing, suitably chosen to
limit the systematic errors from lattice artifacts. Sec.~2 is dedicated to the description of the
general aspects of the  calculation, sec.~3 to its specific details and to the results plus some comments.

\section 2. Theoretical framework

This section is meant to set the notation, define the lattice observables and describe the strategy of the calculation. A recipe for defining
the heavy--light states on a finite volume is then discussed.

\subsection 2.1 General strategy

The calculation of $f_B$ is done on a set of lattices with topology $T\times L^3$ in the Schr\"odinger Functional scheme
 [\ref{SF},\ref{Allscales}], where gauge and fermion fields fulfill periodic boundary conditions along the space directions and Dirichlet boundary 
conditions at the beginning and at the end of the lattice history, and the following set of parameters is used
\equation{
T = 2L,\qquad C=C'=0,\qquad \theta = 0
\enum}

Here $C$ and $C'$ are the boundary gauge fields and $\theta$ is a topological angle which enters into the definition of the Schr\"odinger
Functional. In order to have a safe extrapolation to the continuum, non pertubatively $O(a)$ improved action [\ref{cSWcA}] and operators are used. 
Within this framework, the gauge invariant correlation functions which describe the propagation of a heavy--light pseudoscalar meson from the
low boundary to the bulk and across the two boundaries are
\equation{
\euf_A( x_0 ) = -{L^3\over 2}\langle A_0^I(x) \cS\rangle,\qquad \euf_1 = -{1\over 2}\langle \cS \cS'\rangle
\enum} 

where the operators $A_0^I(x)$, $\cS$ and $\cS'$ interpolate the meson field and are given by
\equation{
A_\mu^I(x) = \psibar_h(x)\gamma_\mu\gamma_5\psi_\ell(x) + {\ ac_A \over 2}\ (\partial_\mu^* + \partial_\mu)\ [\psibar_h(x)\gamma_5\psi_\ell(x)]
\enum}
\equation{\cS = {a^6\over L^3}\sum_{\by,\bz}\ \zetabar_\ell(\by)\gamma_5\zeta_h(\bz),\qquad 
          \cS' = {a^6\over L^3}\sum_{\by',\bz'}\ \zetabar_h(\by')\gamma_5\zeta_\ell(\bz')
\enum}

We take the $O(a)$ improvement coefficient $c_A$, which appears in the axial current, from [\ref{cSWcA}]. The quantum mechanical representation
 of the  correlations functions (2.2) can be found in [\ref{massSF}]. Here we report for convenience their large time behavior, 
referring the reader to that paper for the notation:
\equation{
\euf_A(x_0) = {L^3\over 2}\rho \ \langle 0,0|A_0|0,h\ell\rangle\ \exp\bigl(-x_0M_{h\ell}\bigr)\times\biggl[1+\eta^B_Ae^{-x_0\Delta} +
 \eta^0_Ae^{-(T-x_0)m_G}\biggr]
\enum}
\equation{
\euf_1 = {1\over 2}\rho^2\ \exp\bigl(-TM_{h\ell}\bigr)
\enum}

The state $|0,h\ell\rangle$ represents by definition the lowest heavy--light eigenstate with the quantum numbers of a pseudoscalar, 
and the matrix element $\langle 0,0|A_0|0,h\ell\rangle$ is related to the heavy--light meson decay constant through the relation
\equation{
\hat Z_A\langle 0,0|A_0|0,h\ell\rangle = f_{h\ell}M_{h\ell}(2M_{h\ell}L^3)^{-1/2}
\enum}

with the improved axial current renormalisation constant $\hat Z_A$ given by
\equation{
\hat Z_A = Z_A(1+b_Aa\hat m),\qquad \hat m = {m_\ell + m_h\over 2}
\enum}

The axial current renormalization constant $Z_A$ has been computed non perturbatively for the $O(a)$ improved theory in [\ref{axial}]. For
what concerns the improvement coefficient $b_A$, we know 
from ref.~[\ref{gupta}] that already at $\beta = 6.4$ the discrepancy between the one loop calculation and a non-perturbative
one is of the order of few percent. For this reason we assume for $b_A$ the perturbative value quoted in [\ref{bA}] introducing, 
in our final result, a systematic error that is below 1\%. 
For large times, eq.~(2.5) could be used in order to determine $M_{h\ell}$, while the decay constant
 $f_{h\ell}$ could be extracted from the ratio
\equation{
\hat Z_A{\euf_A(x_0)\over \euf_1} \simeq {1\over2}f_{h\ell}(M_{h\ell}L^3)^{1/2}e^{-(x_0-T/2)M_{h\ell}}
\biggl[1+\eta_A^Be^{-x_0\Delta} + \eta_A^0e^{-(T-x_0)m_G}\biggr]
\enum} 

In a finite size time extension, the asymptotic expansions (2.5--2.6) are in general not valid, and it's impossible to disentangle 
the lowest state contribution to the correlation functions from the excitations due to the higher states. 
We define masses and decay constants at a value of $x_0$ which is a fixed fraction of the
total time extent of the lattice, e.g. $x_0 = T/2$. This affects the finite size observables with spurious
contributions from the  excited states, if $T$ is not sufficiently large, but the recursive procedure
 which connects the small volumes to the large ones through the step scaling function, also connects small times
 to large times, and the final result comes out
to be projected onto the fundamental state. For this reason we choose to measure masses and decay constants at $x_0=L\equiv T/2$, through the equation
\equation{
f_{h\ell}(L) = {2\ \hat Z_A\over \sqrt{M_{h\ell}L^3}}{\euf_A(L)\over \sqrt{\euf_1}}
\enum}

\vskip .3cm
All the simulations are done in the quenched theory, and the connection to physical units is done fixing the scale $r_0$ [\ref{rzero}] at
$r_0 = 0.5$ fm. This is a convenient choice for quenched QCD, where the ratio $r_0/a$ has been computed for a wide range of $\beta$'s with
high precision [\ref{scaleone},\ref{scaletwo}] and can be connected at
higher values of $\beta$ with the behaviour expected from asymptotic
freedom (see ref.~[\ref{toappeartwo}] for details).

\subsection 2.2 Heavy--light mesons on a finite volume

An important aspect of the calculation is the tag of the heavy--light meson states at finite volume. This is required for both the computation
of the decay constant $f_{h\ell}(L_0)$ on the smallest volume and for the evolution steps to the larger ones. The identification has to be done in terms
of a physical quantity which is independent from the volume and we choose it to be the Renormalization Group Invariant (RGI) quark mass. The recipe we
 propose is the following: first of all, we monitor the quark masses via the axial Ward identity
\equation{
\partial_\mu A_{\mu}^I(x) = ( m_{WI}^\ell +  m_{WI}^h) P(x), \qquad P(x) =  \psibar_h(x)\gamma_5\psi_\ell(x)
\enum}

Then, we connect the Ward Identity masses $m_{WI}$ to the renormalization group invariant (RGI) masses $M$ through the relation
\equation{
M = \hat Z_M(g_0)\ m_{WI}(g_0) \equiv Z_M(g_0){(1+b_Aam_{\rm q})\over (1+b_Pam_{\rm q})}\ m_{WI}(g_0)
\enum}

where
\equation{
Z_M(g_0) = {M\over \bar m(\mu)}{Z_A(g_0)\over Z_P(g_0,\mu)},\qquad \mu = {1\over L}
\enum}

The renormalization constants $Z_P$ and $Z_M$ have been determined in [\ref{rgimass}]. Regarding to the improvement coefficients $b_A$ and $b_P$, 
for a sufficiently small subtracted quark mass in lattice units $am_{\rm q}$, what is really needed is the difference  $b_A - b_P$, which is known 
non perturbatively from [\ref{bAbP}]. The identification of the heavy--light meson states on a finite volume is done by expressing all the observables 
as functions of the RGI masses and extracting their values at the physical points computed in literature and reported in Tab.~1.
\vskip -0.6cm
\midinsert
\newdimen\digitwidth
\setbox0=\hbox{\rm 0}
\digitwidth=\wd0
\catcode`@=\active
\def@{\kern\digitwidth}
\tablecaption{RGI masses}
\vskip 1.5ex
$$\vbox{\settabs\+x&xxxxxx&xxxxx&xxxxxx&xxxxx&xxxxxxxxxx&xxxxx\cr
\thicktablerule
\vskip1ex
                \+& \hfill $f$ \hfill
                 && \hfill $M_f\ ({\rm GeV})$\hfill
                 && \hfill ${\rm ref.}$\hfill
                 &&  \cr
\vskip1.0ex
\thintablerule
\vskip1.5ex
  \+& \hfill $s$ \hfill
  &&  \hfill $0.138(6)$ \hfill
  &&  \hfill [\ref{Ms}] \hfill
  &\cr
  \+& \hfill $c$ \hfill
  &&  \hfill $1.684(64)$ \hfill
  &&  \hfill [\ref{Mc}] \hfill
  &\cr
  \+& \hfill $b$ \hfill
  &&  \hfill $7.01(3)(10)$ \hfill
  &&  \hfill [\ref{Mb}] \hfill
  &\cr
\vskip1ex
\thicktablerule
}$$
\endinsert
\vskip 0.8cm
\subsection 2.3 Convergence of the inversion algorithm for the heavy quark propagator

Propagators are computed using the BiCGStab algorithm with SSOR preconditioner for the inversion of the Dirac--Wilson operator 
plus the clover term [\ref{SSOR}]. The inversion is done with 32bit arithmetic both for the light and the heavy quark, and in principle one could 
ask weather rounding effects are present for the heavy quark case, where the exponential decay of the propagator is quite steep. As it will be
explained in sec.~3, the parameters of all the simulations have been chosen so to keep an upper bound on the bare heavy quark mass such that 
$am_h\lesssim 1/3$. In this situation the inversion is expected to be safe. Nevertheless, the quality of the inversion can be monitored with two simple
checks. The first one follows from the observation that the exponential decay of a heavy quark propagator in a non trivial fixed gauge background
has small fluctuations around the tree--level path due to the heaviness of the quark, and the rounding effects can be reliably monitored with an
analytic comparison at tree--level. The second one is suggested by the fact that the quarkonia states $\bar h h$ can be accomodeted on a small
volume without sensible finite size effects, and the rounding on the propagator can be monitored looking at the mass spectrum for these heavy--heavy
states. All the simulations we made did pass the two checks.  

\section 3. The specific calculation and the results

The results are obtained at finite lattice spacing.
The size of the smallest volume follows from the decision of making our estimate for the  finite size $f_B$ on a $48 \times 24^3$ lattice
with a cutoff of about $a_0^{-1}\simeq 12\ {\rm GeV}$. The value of
the bare coupling for this lattice spacing has been obtained from a
fit in  ref.~[\ref{toappeartwo}].
The procedure fixes $\beta(a_0) = 7.3$ and the physical volume $L_0 = 0.4$ {\rm fm}.  
On this lattice, we simulate heavy quark masses up to 0.3 in lattice units, corresponding
to bare physical masses slightly above $4$~GeV.
Indeed, as a general caution against large lattice artifacts, at all $\beta$ values we take
the maximum heavy quark mass in lattice units of the order of 0.3.
The first $\Sigma$ (we distinguish between the continuum step function $\sigma$ and
the one at finite lattice spacing $\Sigma$) goes from  the volume of $0.4$ {\rm fm} to the one of $0.8$~{\rm fm}.
In terms of lattice points, we go from 12 to 24, and we have to match
the starting volume of $0.4$ {\rm fm} with a resolution which is half of the one used for
a correct estimate of the bottom quark propagation. According to our caveat,
it follows that the maximum  bare  quark mass that we can achieve is correspondingly
halved, i.e. of about a couple of GeV at a bare coupling $\beta = 6.737$. 
We make a further iteration with a second $\Sigma$ going from
$0.8$~{\rm fm} to $1.6$~{\rm fm}, where our investigation of heavy
quark masses stops at the order of the charm quark mass. The corresponding bare coupling is $\beta = 6.211$.
The finite volume effects for this second evolution step are small enough
to make the neglection of the residual volume effects a safe assumption, that however can be tested
explicitly.
\vskip 0.3cm
The plots in Figs.~1 and~2 show the dependence of $\Sigma$  upon the {\it heavy} RGI quark mass $M_{RGI}^h$ for the two volume jumps and provide
evidence for a plateau of insensitivity to heavy quark masses: the first three sets of data represent the measured values of 
$\Sigma$ at fixed values of the {\it light} quark mass $M_{RGI}^\ell$, and the other two have been obtained from a linear
extrapolation in $M_{RGI}^\ell$ to the down and strange RGI quark masses reported in Tab.~1. The detail plots show a fit, in the region
of large quark masses only, to the $M^h_{RGI}$ dependence of $\Sigma$ reported here against $1/M^h_{RGI}$ and confirm the validity of
the expansion, given the small slope of the $1/M^h_{RGI}$ correction.  The dependence upon the {\it light} quark mass for fixed heavy quark masses
is given in Figs.~3 and~4. These figures are the core of this paper and support the procedure proposed.
\midinsert
\vbox{
\vskip0.0cm
\epsfxsize=12.5cm\hskip1.2cm\epsfbox{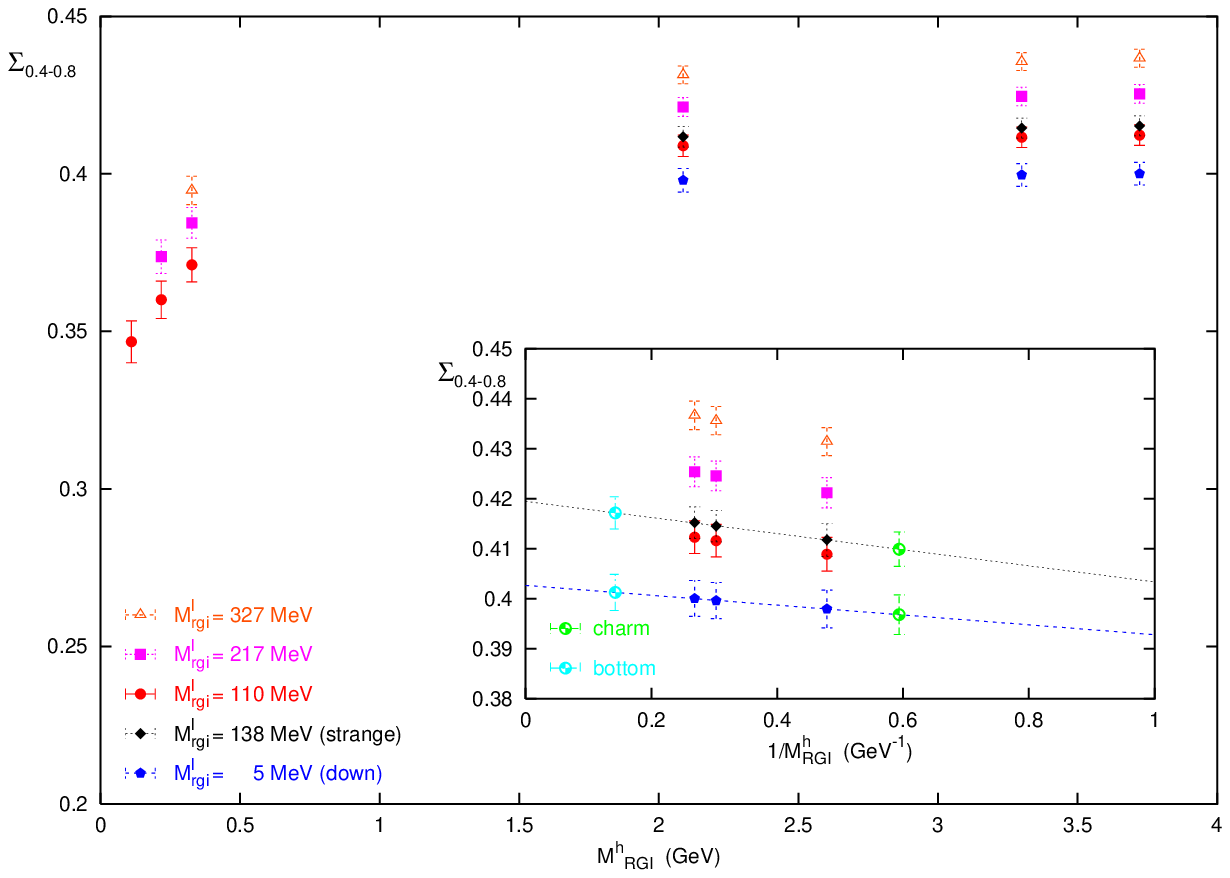}
\vskip-0.1cm
\centerline{
\figurecaption{%
\eightrm Step scaling function $\Sigma_{0.4-0.8}$ for the evolution of $f_{h\ell}$ from $0.4$ {\rm fm} to $0.8$ {\rm fm} at $\beta = 6.737$.}}
\vskip0.7cm
\vskip0.0cm
\epsfxsize=12.5cm\hskip1.2cm\epsfbox{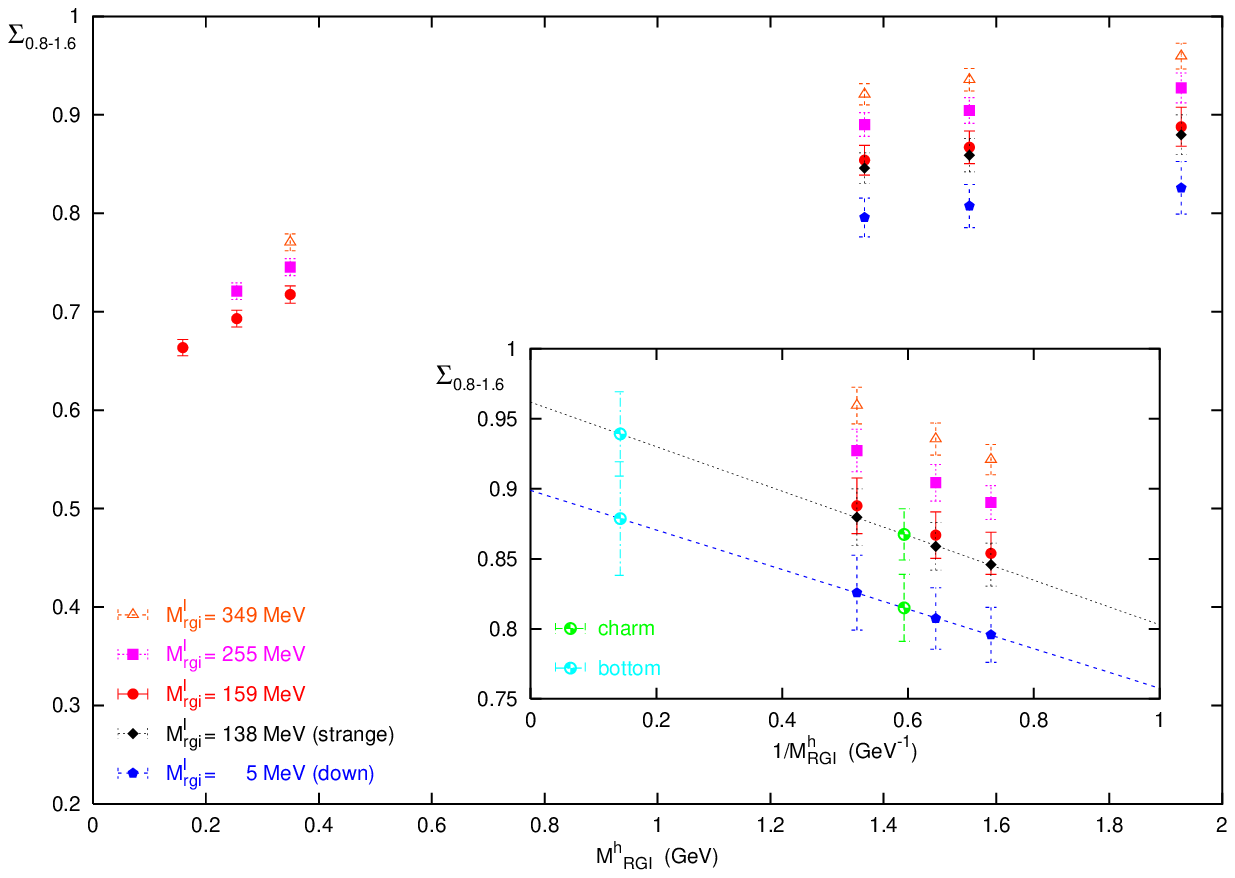}
\vskip-0.1cm
\centerline{
\figurecaption{%
\eightrm Step scaling function $\Sigma_{0.8-1.6}$ for the evolution of $f_{h\ell}$ from $0.8$ {\rm fm} to $1.6$ {\rm fm} at $\beta = 6.211$.}}
\vskip0.0cm
}
\endinsert
\topinsert
\vbox{
\vskip0.0cm
\epsfxsize=12.5cm\hskip1.2cm\epsfbox{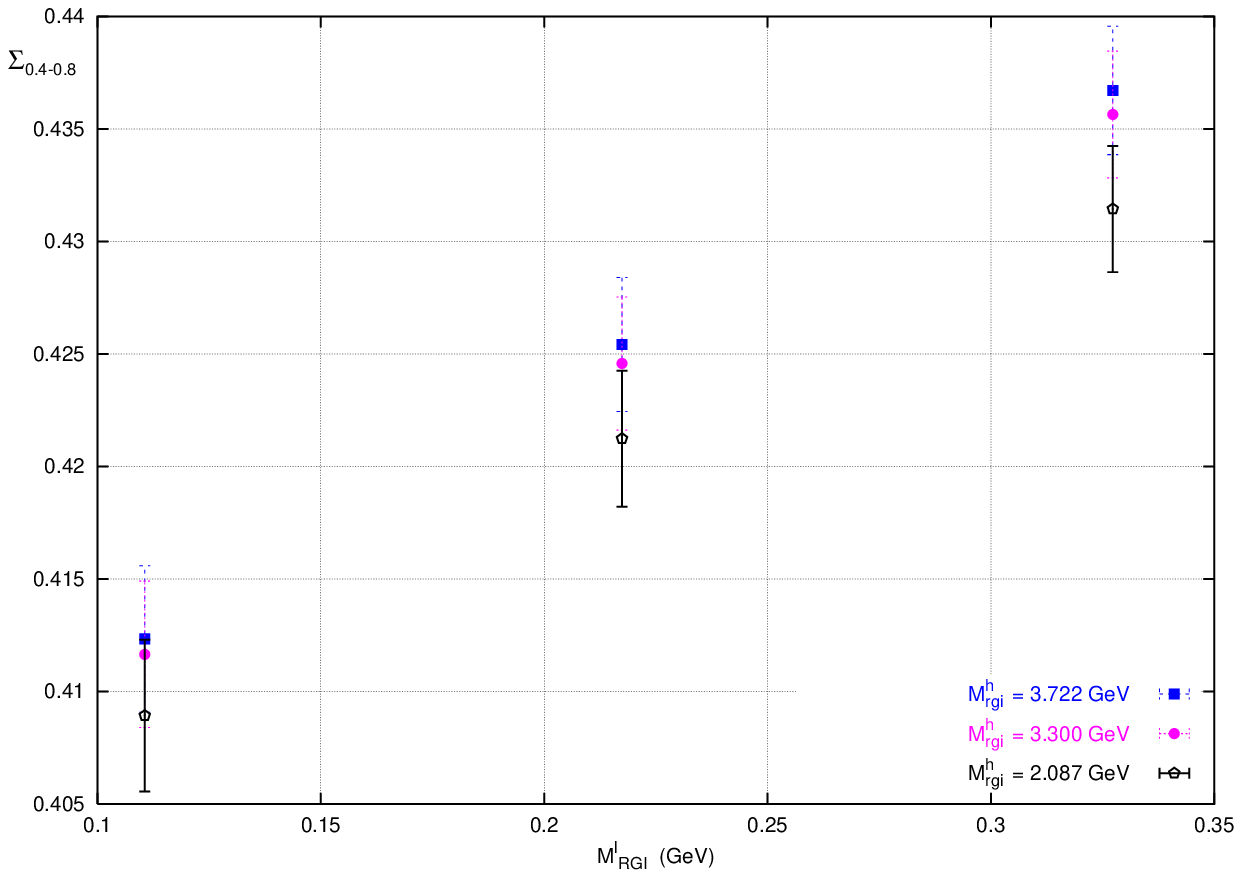}
\vskip0.0cm
\centerline{
\figurecaption{%
\eightrm Dependence of $\Sigma_{0.4-0.8}$ at $\beta = 6.737$ on the light quark RGI mass at fixed heavy quark mass.}}
\vskip0.7cm
\epsfxsize=12.5cm\hskip1.2cm\epsfbox{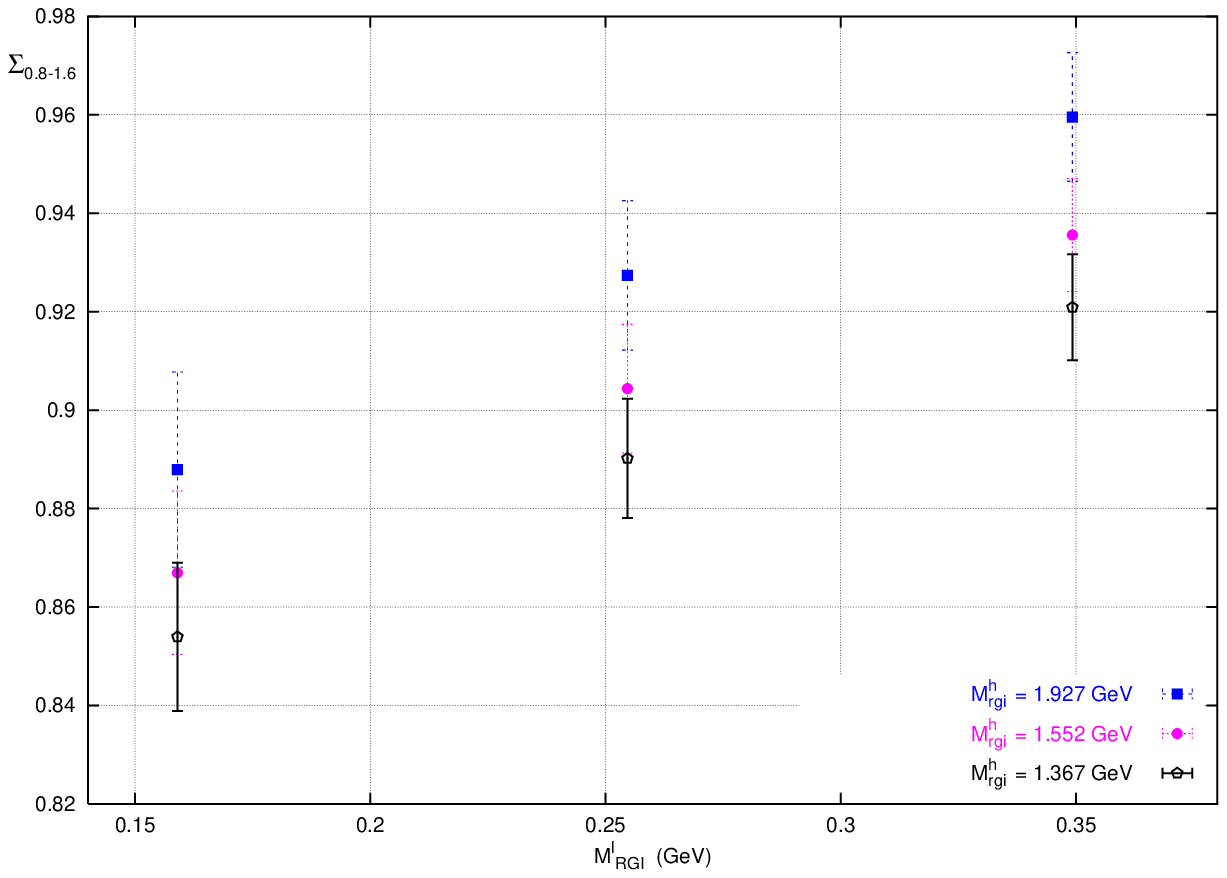}
\vskip0.0cm
\centerline{
\figurecaption{%
\eightrm Dependence of $\Sigma_{0.8-1.6}$ at $\beta = 6.211$ on the light quark RGI mass at fixed heavy quark mass.}}
\vskip0.2cm
}
\endinsert
\vskip 0.5cm

The statistical errors are computed by a jacknife method. More specifically, the errors for the values of $\Sigma$ at 
the physical points, i.e. at the values of the charm and of the bottom RGI quark masses, 
are obtained by making two independent jacknife fits to the $1/M^{h}_{RGI}$ dependence to the
 numerator and to the denominator of eq.~(1.1).  In this way, we avoid dealing with results at various mass 
values correlated by the same set of background gauge
configurations. The final error on the $\Sigma$'s is obtained by combining quadratically the
relative statistical errors resulting from the two jacknife fits.

The finite size value  of $f_B$ is obtained by a calculation on the highest resolution lattice.
The RGI bottom quark mass, according to the previous section, is obtained from the equation
\equation{
M = \hat Z_M(g_0)\ m_{WI}(g_0)
\enum}

In order to obtain the renormalisation constant $Z_M(g_0)$  at $\beta = 7.3$ and $\beta = 6.737$, we have used 
a safe interpolation of the pseudoscalar renormalisation constant $Z_P(g_0,\mu)$ at a value of $\mu$ 
three times the reference value used in eq.~(6.8) of ref.~[\ref{rgimass}]. The value for $f_B$ that we obtain is
\equation{
f_B(\ 0.4{\rm \ fm}\ ) = \fBsmall \ {\rm MeV}
\enum}

By using the values of $\Sigma$ for the $b$ quark at constant RGI mass marked in figures 3 and 4
\equation{
\Sigma_{0.4-0.8}^{bd} = 0.401(4),\qquad
\Sigma_{0.8-1.6}^{bd} = 0.88(4)
\enum}

we obtain our estimate of $f_B$ on the large volume:
\equation{
f_B^{phys} \equiv f_B(\ 0.4{\ \rm fm}\ )\cdot \Sigma_{0.4-0.8}^{bd} \cdot \Sigma_{0.8-1.6}^{bd} = \fBstat \ {\rm MeV}
\enum}

where the error quoted in the previous equation is statistical only. 

The {\it systematic errors} can be partly ascribed to specific approximations used in the present computation that 
can be eventually removed, and partly to the uncertainty in the extrapolation in the heavy quark mass of
finite size effects, inherent to the method proposed.

To the first class belong the errors related to a finite lattice spacing
both for the step scaling function and for the finite size decay constant.
The former introduce a new dimensionful variable   $aL$ in the finite lattice spacing
step scaling function.
The use of an non-perturbatively improved action, but for surface counterterms that
are evaluated in perturbation theory, makes the error at most of 
$O( \alpha^{\rm lattice} \ a/L )$  or $O( \alpha^{\rm lattice} \ a
\Lambda_{QCD} )$ i.e. of  a few percent.
Notice that lattice artefacts related to the heavy quark mass alone
cancel in the ratio defining the step scaling function at diffenet
lattice sizes, but at the same values of the cutoff and of the quark mass.

Preliminary results from a calculation where the continuum limit is
estimated indicate that such is the case~[\ref{workinprogress}].

We estimate an uncertainty for each finite lattice spacing step scaling function of about $2$\%.
The lattice artifacts of order $(aM_h)^2$ remain in the determination  of the
finite size $f_B$. With our restriction on the maximum value of the
heavy quark mass in lattice units, 
we can limit this uncertainty to less than
$10$\%.
The overall effect of finite lattice spacing on the quantity in eq.~(1.2) is not expected to exceed $12-13$\%.

A second source of uncertainty derives from our estimate of the
lattice spacing at large $\beta$ obtained from the asymptotic freedom fit of ref.~[\ref{toappeartwo}].
This produces a variation of $f_B$ on the small volume in an
indirect way. 
If the lattice spacing, say, is larger than estimated, the value of
$f_B$ translated in physical units 
is accordingly smaller. However, in such a case, the volume used is
larger than the expected $0.4$ {\rm fm}: one must repeat the
calculation at higher beta, on the matched physical volume, where
the value of $f_B$ is {\it higher}, because of finite size effects.
The variation induced by an error in the lattice spacing
depends upon the finite size dynamics. In order to estimate it,
one has to get the error on $\beta$ for a fixed lattice spacing and
perform test simulations within the error range.
From ref.~[\ref{toappeartwo}] this amounts to a $0.2$\% error.  We have performed
simulations  at $\beta=7.2$ and of $\beta=7.3$, a variation range ten times
bigger than the error quoted, and obtained a variation of
$f_B$ on the small volume much below our statistical error that must then considered a
generous upper bound on this effect.
An independent estimate could come from bottomonium spectroscopy on
a finite volume that would also supply an addidtional estimate of the bottom quark mass.

A minor source of uncertainty, negligible  and anyway removable, come from our derivation  of the 
renormalisation constant that determines the RGI invariant mass from the ward identity mass at the highest $\beta$.

To the first class finally belong the residual finite volume effects beyond the lattice size of $1.6$~{\rm fm} that
we have simulated. This is a volume considered safe for numerical simulations of light quark spectroscopy
and we do not expect residual corrections. As already mentioned, a specific test can anyway be made by 
calculating the next step function.

To the second class of errors belong the ones deriving from the extrapolation of the step function to values
of the heavy quark mass higher than the ones simulated, i.e. the validity of the asymptotic expansion
of eq.~(1.2). This can be partly eliminated by running more quark mass values and by constraining further the fit.
Our data lign on a straight line very well. We estimate an error from a parabolic fit through our three points of about $1$\%.

The overall error on the number $f_B$  coming from the removable systematic uncertainties is of about $13$\% and
of at most $2-3$\% from the ones deriving from the unavoidable extrapolation in the heavy quark mass, 
leading to a global uncertainty of about $25$~{\rm MeV} of which about $20$ are removable while $5$ stay with the method:
\equation{
f_B^{phys} = \fB \ {\rm MeV}
\enum}

As a check of the whole procedure, we have made the estimate for the charm quark case,
whose RGI mass from ref.~[\ref{Mc}] provides a very good fit to the spectroscopy 
on the largest volume at $\beta = 6.211$:
\equation {
M_D = 1.814(6)\ {\rm GeV}, \qquad M_{\eta_c} = 2.881(2)\ {\rm GeV}
\enum}

For the charm quark, the value of the decay constant $f_D$ coming from the finite size procedure can be compared with the value
obtained directly on the large volume $L = 1.6$ {\rm fm} at $\beta = 6.211$. The finite size decay constant is
\equation{
f_D( \  0.4\ {\rm fm}\ ) = \fDsmall \ {\rm MeV}
\enum} 

while the step scaling functions are
\equation{
\Sigma_{0.4-0.8}^{cd} = 0.397(4), \qquad
\Sigma_{0.8-1.6}^{cd} = 0.81(2)
\enum}

The comparison between the two determinations is 
\equation{\left\{\eqalign{
& f_D^{phys} = f_D(\ 0.4\ {\rm fm}\ )\cdot \Sigma_{0.4-0.8}^{cd} \cdot \Sigma_{0.8-1.6}^{cd} = \fD \ {\rm MeV} \cr
& f_D(\ 1.6\ {\rm fm} \ )_{\beta = 6.211} = 208(6)\ {\rm MeV} } \right .
\enum}

The agreement of the two numbers gives us confidence on the reliability of the $f_B$ result.
We have also extracted the value of $f_{B_s}$ by the same procedure, with the finite size decay constant and the step scaling functions for the two
jumps given by
\equation{
f_{B_s}(\ 0.4\ {\rm fm}\ ) = \fBssmall \ {\rm MeV}, \qquad 
\Sigma_{0.4-0.8}^{bs} = 0.417(3),\qquad
\Sigma_{0.8-1.6}^{bs} = 0.94(3)
\enum}

The infinite volume result for this decay constant is
\equation{
f_{B_s}^{phys} = \fBs \ {\rm MeV}
\enum}

The light quark mass dependence of the volume effects is larger when the volume is large enough
to resolve the difference between a strange and light quarks.

Other results coming from our calculations are:
\equation{
f_{B_s}^{phys} / f_B^{phys} = \fBsd
\enum}

and
\equation{
f_{D_s}^{phys} / f_D^{phys} = \fDsd
\enum}

where the first error is statistical and the second comes from the uncertainity
in the light quark extrapolations.

Finally, we have explored the validity of the phenomenological ansatz of eq.~(1.3) for $\Sigma$ and 
Fig.~7 shows $\Sigma_{0.4-0.8}$ as a function of the reduced mass. A reasonable scaling is obtained setting
$ \mu_i = m_i + M_0 $, where $M_0 = 0.5$ GeV, not far from $\Lambda_{QCD}$.
\midinsert
\vbox{
\vskip0.0cm
\epsfxsize=12.5cm\hskip1.2cm\epsfbox{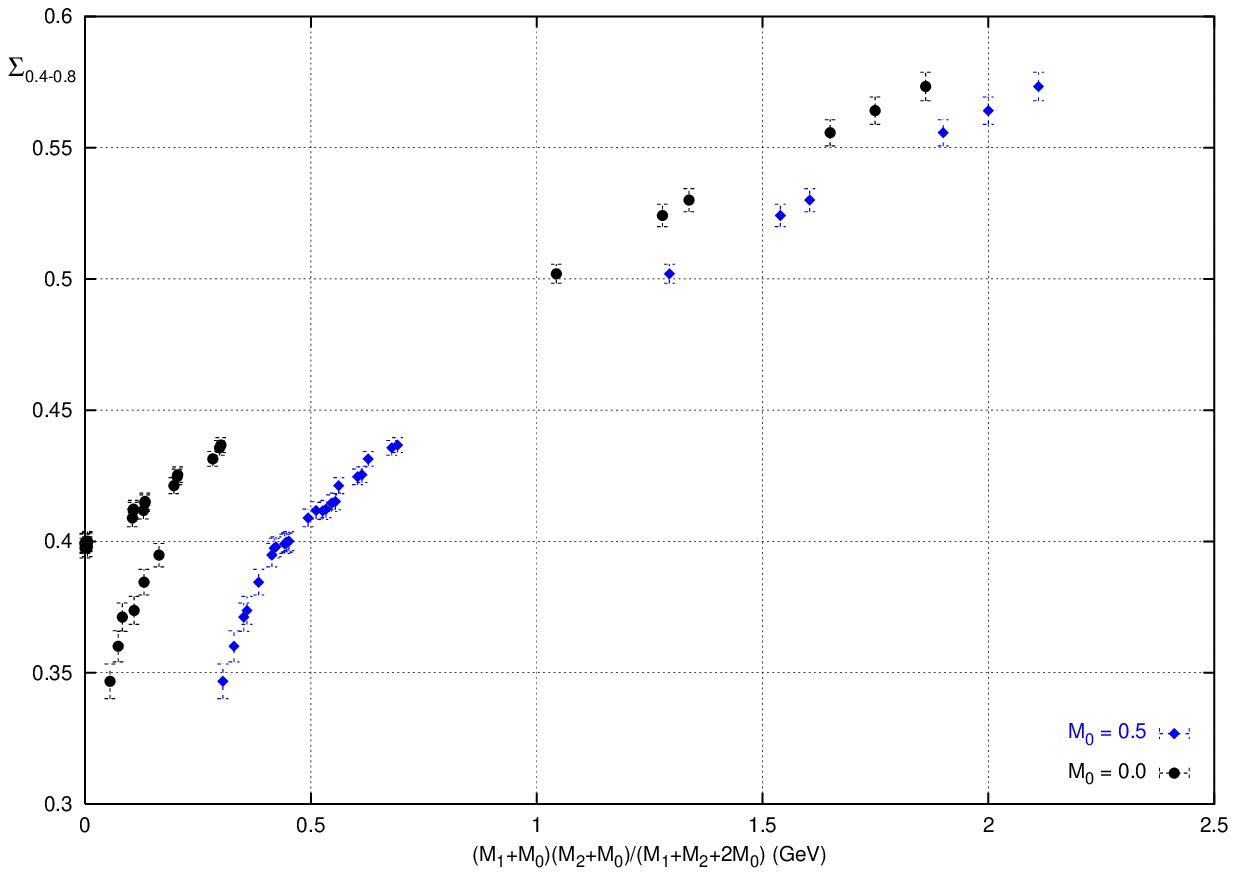}
\vskip-0.1cm
\centerline{
\figurecaption{%
\eightrm Step scaling function $\Sigma_{0.4-0.8}$ as a function of the reduced mass. Here $M_i$ represents the RGI mass.}}
\vskip0.7cm
}
\endinsert
We stress again that an estimate of the bottom quark mass and of the lattice spacing at the smallest volume 
can also be made independently by a fit to the bottomonium
spectroscopy on the smallest size, highest resolution lattice, under the hypothesis
that such flavourless states have a much smaller radius than the flavoured mesons
and do not suffer from finite size effects.
The difficulty of disentangling the contribution of excited states 
that, measured on the scale of the heavy state, are almost degenerate in mass, could be overcome
by optimizing the SF source to obtain the best projection on the fundamental state.

The major systematic effects present in this exploratory calculation can be eliminated, while
the unavoidable errors due to the extrapolations in the heavy quark mass are strongly
suppressed by the manifest insensitivity of finite size effects to the heavy quark mass,
which is the main result of this paper.
All the steps of the calculation always deal with physical quantities properly
renormalized in massless lattice QCD.

The method proposed can be generalized to problems characterised by two very different mass scales,
if the decoupling of the large mass scale from the low scales of non-perturbative QCD dynamics holds true.
This appears to be the case in the example discussed and is somehow supported by the wide success
of the predictions of perturbative QCD calculations for hard processes that are insensitive to the dressing
mechanism of quarks and gluons into standard hadronic final states.
\vskip .8cm
{\bf Acknowledgements} 
\vskip 0.3cm
We thank M.~L\"uscher and N.~Cabibbo for useful discussions. This work has been partially supported by the European Community 
under the grant HPRN--CT--2000--00145 Hadrons/Lattice QCD. 

\beginbibliography


\bibitem{litone}
S.~M.~Ryan, \NPBproc 106 2002 86--97



\bibitem{SF}
M.~L\"uscher et al., \NPB B384 1992 168--228

\bibitem{Allscales}
K.~Jansen et al., \PLB B372 1996 275--282

\bibitem{cSWcA}
M.~L\"uscher et al., \NPB B491 1997 323--343

\bibitem{massSF}
M.~Guagnelli et al., \NPB B560 1999 465--481

\bibitem{axial}
M.~L\"uscher et al., \NPB B491 1997 344--364

\bibitem{gupta}
T.~Bhattacharya et al., \NPBproc 106 2002 789--791

\bibitem{bA}
S.~Sint, P.~Weisz, \NPB B502 1997 251--268

\bibitem{rzero}
R.~Sommer, \NPB B411 1994 839--854

\bibitem{scaleone}
M.~Guagnelli et al., \NPB B535 1998 389--402

\bibitem{scaletwo}
S.~Necco and R.~Sommer, \NPB B622 2002 328--346


\bibitem{rgimass}
S.~Capitani et al., \NPB B544 1999 669--698

\bibitem{bAbP}
M.~Guagnelli et al., \NPB B595 2001 44--62 

\bibitem{Ms}
J.~Garden et al., \NPB B571 2000 237--256

\bibitem{Mc}
J.~Rolf, S.~Sint, \NPBproc 106 2002 239--241

\bibitem{Mb}
J.~Heitger, R.~Sommer, \NPBproc 106 2002 358--360


\bibitem{SSOR}
M.~Guagnelli, J.~Heiger, \CPC 130 2000 12


\bibitem{workinprogress}
M.~Guagnelli, F.~Palombi, R.~Petronzio, N.~Tantalo, work in preparation

\bibitem{toappeartwo}
M.~Guagnelli, R.~Petronzio, N.~Tantalo, hep-lat/02???????

\endbibliography
\vfill\eject
\end